\documentclass[11pt]{article}
\usepackage[preprint]{acl}
\usepackage{times}
\usepackage{latexsym}
\usepackage[T1]{fontenc}
\usepackage[utf8]{inputenc}
\usepackage{microtype}
\usepackage{inconsolata}
\usepackage{graphicx}
\usepackage{cuted}
\usepackage{multicol}
\usepackage{caption}
\usepackage{dblfloatfix} 
\usepackage[ruled,boxed,linesnumbered]{algorithm2e}
\SetInd{0.5em}{1em}
\SetKwInOut{Input}{Input}
\SetKwInOut{Output}{Output}
\SetKwFunction{FindWitness}{FindWitness}
\SetKwFunction{Refine}{Refine}
\SetKwFunction{Gen}{Gen}
\SetKwFunction{Analyze}{Analyze}
\SetKwFunction{Oracle}{Oracle}
\SetKwFunction{Improve}{Improve}
\SetKwProg{Fn}{Function}{:}{end}

\SetCommentSty{algcommentstyle}
\SetKwComment{tcp}{// }{}

\usepackage{float}
\usepackage{newfloat}
\usepackage{listings}
\DeclareCaptionStyle{ruled}{labelfont=normalfont,labelsep=colon,strut=off} % DO NOT CHANGE THIS
\floatstyle{plain}
\newfloat{listing}{tb}{lst}{}
\floatname{listing}{Listing}

\usepackage{booktabs}

\usepackage{color}
\usepackage{amsmath,amsopn}
\usepackage{amssymb}
\usepackage[dvipsnames]{xcolor}
\usepackage[labelformat=simple]{subcaption}

\usepackage{endnotes,microtype,xspace,fancyvrb,multirow}
\usepackage{booktabs}
\usepackage{array,underscore,relsize}
\usepackage[T1]{fontenc}
\usepackage{enumitem}
\renewcommand{\arraystretch}{1.1} % set the spaces between table rows. look nicer :-)
\usepackage{tabularx}
\usepackage{colortbl}

\usepackage[most]{tcolorbox}

\usepackage{fp}
\usepackage{siunitx}
\usepackage{diagbox}

\usepackage{listings}
\definecolor{lightgray}{rgb}{0.91, 0.91, 0.91}
\definecolor{lightblue}{rgb}{0.8627, 0.9176, 0.9686} % consistent with figure blue

\usepackage{verbatim}

\usepackage[capitalise,noabbrev]{cleveref}
\usepackage{pifont}% http://ctan.org/pkg/pifont

\usepackage{xurl}

\fvset{fontsize=\footnotesize,xleftmargin=8pt,numbers=left,numbersep=5pt}

\providecommand{\autoref}[1]{\cref{#1}}
\crefname{lstlisting}{Listing}{Listings}
\crefname{listing}{Listing}{Listings}
\crefname{algorithm}{Algorithm}{Algorithms}
\Crefname{algorithm}{Algorithm}{Algorithms}
\crefname{algocf}{Algorithm}{Algorithms}
\Crefname{algocf}{Algorithm}{Algorithms}

\crefname{section}{\S}{\S\S}
\Crefname{section}{\S}{\S\S}
\crefname{subsection}{\S}{\S\S}
\Crefname{subsection}{\S}{\S\S}

\crefname{equation}{Formula}{Formulas}
\Crefname{equation}{Formula}{Formulas}

\input{glyphtounicode}
\newcommand{\squishlist}{
\begin{itemize}[noitemsep,nolistsep,leftmargin=10pt]
}
\newcommand{\squishend}{
  \end{itemize}
}

\usepackage{tikz}
\newcommand*\WC[1]{%
\begin{tikzpicture}[baseline=(C.base)]
\node[draw,circle,inner sep=0.2pt](C) {#1};
\end{tikzpicture}}

\usepackage{xstring}

\newtcolorbox{mybox}[3][float=ht]
{
  colback=#2!5!white,
  colbacktitle=#2!15!white,
  boxrule=0.25mm, 
  top=0pt, bottom=0pt, left=0pt, right=0pt,
  coltitle=black,
  title    = {#3},
  #1,
}

\newcounter{prompt}
\newlength{\markwidth}
\definecolor{codeqlKeyword}{RGB}{0,92,197}
\definecolor{codeqlType}{RGB}{0,128,0}
\definecolor{codeqlAnnotation}{RGB}{111,66,193}
\definecolor{codeqlComment}{RGB}{106,115,125}
\definecolor{codeqlString}{RGB}{163,21,21}

\lstdefinelanguage{CodeQL}{
  sensitive=true,
  keywords={
    and,any,as,asc,avg,by,class,concat,count,desc,else,exists,
    extends,forall,forex,from,if,implies,import,in,instanceof,
    max,min,module,newtype,none,not,or,order,predicate,rank,
    result,select,strictconcat,strictcount,strictsum,sum,super,
    then,this,unique,where,
    abstract,additional,bindingset,cached,deprecated,extensible,
    external,final,language,library,overlay,override,pragma,
    private,query,transient
  },
  morecomment=[l]{//},
  morecomment=[s]{/*}{*/},
  morestring=[b]{"}
}

\lstdefinestyle{CodeQLStyle}{
  language=CodeQL,
  basicstyle=\small\ttfamily,
  keywordstyle=[1]\color{codeqlKeyword}\bfseries,
  keywordstyle=[2]\color{codeqlType}\bfseries,
  keywordstyle=[3]\color{codeqlAnnotation}\bfseries,
  commentstyle=\color{codeqlComment}\itshape,
  stringstyle=\color{codeqlString},
  showstringspaces=false,
  columns=fullflexible,
  keepspaces=true,
  breaklines=true,
  tabsize=2
}

\newcommand{\sys}{\mbox{\texttt{ARQ}}\xspace}

\definecolor{codegray}{rgb}{0.5,0.5,0.5}
\definecolor{codeblue}{rgb}{0.0,0.0,1.0}
\definecolor{codeteal}{rgb}{0.0,0.5,0.5}

\title{ARQ: Agentic CodeQL Query Refinement \\ for C/C++ Vulnerability Detection}
\author{
  Chunyi Wang\thanks{These authors contributed equally.}\textsuperscript{1}\quad
  Yunfei Ke\footnotemark[1]\textsuperscript{1}\quad
  Junfeng Yang\textsuperscript{1}\quad
  Yun-Yun Tsai\textsuperscript{1}\quad
  Penghui Li\textsuperscript{2} \\
  \normalfont\textsuperscript{1}Columbia University \\
  \normalfont\textsuperscript{2}Nanyang Technological University \\
  {\small\normalfont\texttt{\{cw3723,yk3108\}@columbia.edu}\quad
   \texttt{junfeng@cs.columbia.edu}} \\
  {\small\normalfont\texttt{yt2781@columbia.edu}\quad
   \texttt{penghui.li@ntu.edu.sg}}
}

\begin{document}

\maketitle

\begin{abstract}
Query-based static analyzers such as CodeQL encode vulnerability patterns in detection queries. Existing queries can miss vulnerabilities or flag benign code, and broadening them can trade false negatives for false positives. We present \sys, an agentic framework that refines C/C++ CodeQL queries with execution-grounded evidence from synthesized programs. \sys compares a program's instrumented behavior with CodeQL's verdict, then uses validated false-negative and false-positive witnesses, including near-identical program pairs, to guide LLM-based refinement and regression testing.
We evaluate \sys by refining 12 official CodeQL queries with GPT-5.4, Claude-Sonnet-4.6, and Gemini-3.5-flash. On Juliet v1.3 and FormAI v2, refined queries detect up to 119.8\% more true-positive query--sample detections than the original queries while maintaining precision of at least 98.0\%. Moreover, \sys addresses three unresolved CodeQL issues by correctly classifying their supplied reproducers and identifies two previously unreported candidate bugs in \texttt{libpng} and \texttt{zlib}. These results demonstrate that \sys can substantially expand the execution-validated coverage of production CodeQL queries while preserving high precision across diverse benchmarks and real-world cases.
\end{abstract}

\section{Introduction}

Modern software systems such as operating systems, browsers, and embedded systems are usually built with low-level programming languages (e.g., C or C++). Although these languages provide programmers with fine-grained control over system resources, improper resource management often makes programs prone to security vulnerabilities. For instance, buffer-overflows, use-after-free, and double-free~\cite{cppcoreguide,cppptrbad} are among the most severe vulnerability classes in deployed software, with over 33,000 reported CVEs from 2020$\sim$2025~\cite{cvedb}.

To combat these vulnerabilities, static analysis tools such as~\citet{codeqlgithub},~\citet{semgrepdev}, and ~\citet{joernio} use queries written in domain-specific languages (DSL) to encode vulnerable code patterns~\cite{avgustinov2016ql}. These queries are typically written and maintained manually, then used to scan target software for vulnerable code. The GitHub Advanced Security (GHAS) platform uses CodeQL to automatically detect vulnerabilities across millions of repositories.

However, static-analysis queries are difficult to maintain. Refining an existing query safely requires expertise in software systems, software security, vulnerability semantics, and with detailed knowledge of the query DSL~\cite{10.1145/3314221.3314648,li2024iris,li2025automatedstaticvulnerabilitydetection,wang2026qlcoderquerysynthesizerstatic}. A change that recovers one missed vulnerability pattern can also introduce false positives or regressions elsewhere. In practice, reported query bugs remain open for \textbf{6 to 27 months} without a fix, as shown in \cref{s:rq2}. Therefore, this maintenance burden has motivated recent work on automating query construction and refinement with large language models (LLMs).

We present \sys, a framework for targeted refinement of existing C/C++ CodeQL queries using execution-grounded evidence. Given an existing query, \sys proceeds in four main steps: (1) using LLM-guided program synthesis to generate C/C++ programs that expose potential false-positive or false-negative behavior, (2) executing the synthesized programs under a fixed harness to validate their runtime behavior, (3) selecting validated witnesses, including paired programs with contrasting query behavior, and (4) using these witnesses to guide targeted updates to the query. Instead of synthesizing a new query from scratch, our goal is to repair an existing one while addressing observed weaknesses, limiting false-positive growth, and avoiding regressions in prior validated cases.

This setting differs from recent LLM-based approaches to static-analysis query construction and adaptation. QLCoder~\cite{wang2026qlcoderquerysynthesizerstatic}, for example, synthesizes new Java CodeQL queries from CVE metadata, while other systems use taint labels, symbolic validation, project history, vulnerability patterns, or labeled datasets to construct or refine static-analysis queries~\cite{10.1145/3589334.3645530,7961989,li2024iris,li2025automatedstaticvulnerabilitydetection}. In contrast, \sys starts from an existing query and derives its refinement signal from \emph{executable programs}, making the evidence directly verifiable through runtime behavior. For each program, \sys compares the behavior observed during execution with whether the CodeQL query reports the program as vulnerable. Disagreements between these two signals provide validated evidence of false negatives or false positives that can guide query refinement. This execution-grounded feedback allows \sys to identify both missed behaviors and overly broad matches without relying on external vulnerability metadata or manually labeled refinement examples.

This distinction is particularly important for C/C++. IRIS~\cite{li2024iris}, for example, uses an LLM to label taint sources and sinks in Java web applications and detects vulnerabilities by tracking whether tainted data reaches a sink. Such reasoning is well suited to data-flow vulnerabilities such as SQL injection, but many C/C++ memory-safety bugs depend instead on allocation boundaries, object lifetimes, pointer aliasing, and deallocation behavior~\cite{cppcoreguide,cppptrbad}. A pointer may be dereferenced after its target has been freed, or an access may exceed an allocated buffer, even when no tainted data flow is involved. These behaviors therefore require refinement signals that capture concrete execution semantics rather than taint propagation alone. To our knowledge, prior work has not studied execution-grounded refinement of existing C/C++ CodeQL queries.

\noindent We highlight our main contributions:

\begin{enumerate}[itemsep=0pt,topsep=2pt,parsep=0pt,leftmargin=*]
\item We introduce \sys, the first execution-grounded framework that refines existing C/C++ CodeQL queries using runtime-validated witness programs rather than external metadata or static labels.
\item We use contrastive witness pairs to guide targeted updates that recover missed behaviors while limiting false positives and regressions.
\item We assemble a reusable evaluation artifact of CodeQL query-refinement tasks, execution-validated witness programs, regression cases, and per-query results for studying automated static-analysis maintenance.
\item On Juliet and FormAI, \sys improves true-positive query--sample detections by up to 119.8% while maintaining at least 98.0% precision.
\item On real-world cases, \sys addresses 3 unresolved CodeQL issues and finds a developer-confirmed bug in \texttt{zlib} and a candidate bug in \texttt{libpng}.
\end{enumerate}

\section{Preliminaries}
\subsection{Motivating Example: Double-Free Query}
\label{s:background-motivation}

\begin{figure*}[t]
  \centering
  \includegraphics[width=\textwidth]{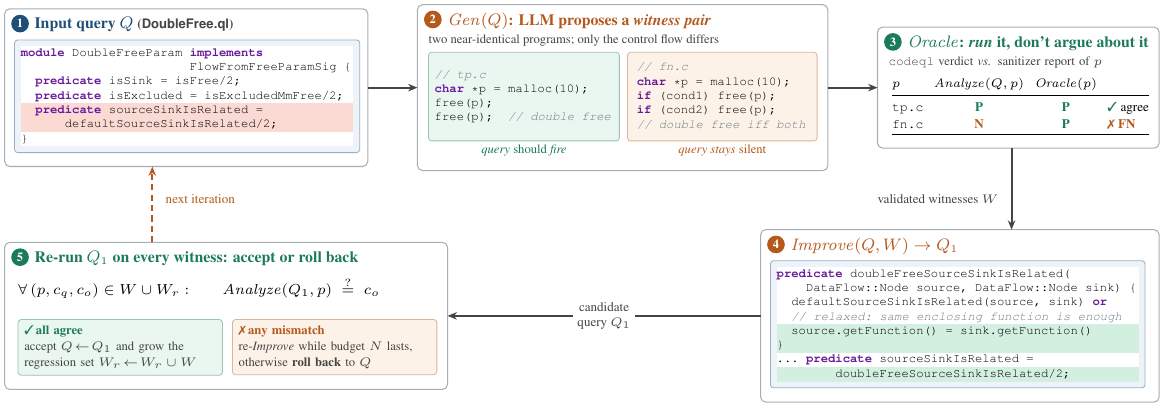}
  \caption{\sys refining \texttt{DoubleFree.ql}. \protect\WC{1} the query under refinement. \protect\WC{2} the LLM proposes a pair of near-identical programs, \texttt{tp.c} and \texttt{fn.c}, that should disagree on the query's verdict. \protect\WC{3} an execution oracle provides evidence about the exercised behavior. \protect\WC{4} guided by this validated witness (\texttt{fn.c}), the LLM proposes a refined query. \protect\WC{5} previously validated witnesses provide regression evidence when assessing the refined query.
  }
  \label{fig:overview}
\end{figure*}

We use \texttt{DoubleFree.ql}, a real query from GitHub's official CodeQL query suite\footnote{\url{https://github.com/github/codeql/blob/d1fed84daf8f3abc59cf9453692b98632932d60f/cpp/ql/src/Critical/DoubleFree.ql}}.
A double-free occurs when a program calls \lstinline[language=C]{free} twice on pointers representing same allocated memory. \texttt{DoubleFree.ql} detects this pattern by tracking each memory allocation as a source and its deallocation calls as sinks, then checking whether two sinks can both execute for the same allocation (step \WC{1} in \cref{fig:overview}).
However, the query defines a custom relation predicate, \lstinline[language=CodeQL]{sourceSinkIsRelated}, that delegates to \lstinline[language=CodeQL]{defaultSourceSinkIsRelated}. The default predicate relates the two sinks only when they occur within the same \lstinline[language=C]{if} branch. This restriction causes a false negative. Two \lstinline[language=C]{free} calls in separate, independent \lstinline[language=C]{if} statements can both execute when both conditions hold, but the query does not relate them because they occur in different branches. This is a query-logic weakness rather than a sanitizer artifact: the same allocation is freed twice, but the query rejects the pair solely because of the branches in which the frees occur.

\subsection{Challenges of Query Refinement}

Automating this refinement presents two challenges. First, query failures often remain latent because they are never reported. A query's silence does not establish that a vulnerability exists, so exposing an observed false negative requires an independent execution signal. Second, broadening a query to recover missed cases can introduce many false positives by admitting benign patterns that the query previously excluded. Neither challenge can be solved by inspecting the query alone. Both require comparing its verdict with the observed behavior of generated programs.

\subsection{\sys: Execution-Grounded Refinement}
Unlike LLM-only reasoning about query correctness, \sys grounds witness validation in program execution. Instrumented execution provides evidence about whether a synthesized program exercises the relevant behavior, while CodeQL provides the corresponding static verdict. Comparing these signals exposes observed query weaknesses. \sys also rechecks candidate refinements against previously validated witnesses, using them as regression tests. This evidence is specific to the supplied program, input, and execution path; it does not establish global program safety or complete vulnerability recall.

\paragraph{Generating and Validating Witnesses}

In step \WC{2}, \sys asks the LLM to generate a pair of candidate programs that may become witnesses after validation. The names \texttt{tp.c} and \texttt{fn.c} denote the verdict each program is meant to elicit, a true positive and a false negative respectively, and the two programs differ only in control flow. \texttt{tp.c} is an observed true positive, since its two \lstinline[language=C]{free} calls are adjacent. \texttt{fn.c} is an observed false negative. Its two \lstinline[language=C]{free} calls instead sit in different \lstinline[language=C]{if} branches, so the relation predicate never holds, even though the same pointer is freed twice whenever both conditions are true. In step \WC{3}, \sys applies \texttt{DoubleFree.ql} and the compiler's sanitizer~\cite{clangsan,gccsan} (compiled with \lstinline[language=bash]{-fsanitize=}) to both programs. The sanitizer reports the exercised double-free in \texttt{tp.c} and \texttt{fn.c}, while \texttt{DoubleFree.ql} does not flag \texttt{fn.c}. This execution/query mismatch validates \texttt{fn.c} as a witness for an observed false negative; it does not establish global program safety or vulnerability beyond the exercised path.

\paragraph{Refining the Query}
Guided by the confirmed witness from step \WC{3}, step \WC{4} has the LLM refine \texttt{DoubleFree.ql} by relaxing \lstinline[language=CodeQL]{sourceSinkIsRelated} to also allow two frees within the same enclosing function. In step \WC{5}, \sys re-validates the refined query against previously confirmed witnesses, including \texttt{tp.c} and \texttt{fn.c}, to assess whether the update introduces a regression. This concludes one iteration of refinement. \sys repeats this loop across multiple iterations, alternating between false-negative and false-positive witnesses, until the refinement budget is exhausted.

\sys alternates between generating and validating contrastive witnesses and using those witnesses to guide query refinement. It repeats this process until the iteration budget is exhausted or no validated witness can be generated. Section 3 formalizes these steps through four components, $Gen$, $Oracle$, $Analyze$, and $Improve$.

\section{\sys Framework}
We formalize the problem as follows. Given an initial query $Q_0$, \sys aims to produce a refined query $Q_f$ through at most $T$ iterations. Each iteration uses a budget $N$ to generate and validate witnesses, propose a query update, and check that update against previously validated witnesses.
During refinement, $Gen(\cdot)$ generates programs that try to expose weaknesses in the current query. $Analyze(\cdot)$ and $Oracle(\cdot)$ validate the generated witnesses, and $Improve(\cdot)$ proposes a candidate query from the validated cases. A candidate is accepted only when it passes current and regression witnesses. \cref{alg:iteration} summarizes this workflow, while the following subsections formally define the four components and their acceptance conditions.

\subsection{Generating and Validating Witnesses}
\label{s:design:witness}
We define an \emph{observed query failure} as a false positive or false negative that the query commits against a specific program $p$ under a fixed execution harness
. The generator $Gen(Q)$ returns candidate programs for the current query. For each executable program $p$, $Analyze(Q,p)$ and $Oracle(p)$ return classifications in $\{P,N\}$. Here, $P$ means that the relevant instrumented vulnerability is observed during execution, and $N$ means that no such error is observed
.
\paragraph{\boldmath$Analyze(Q,p)$} It returns the query's classification
$
c_q = Analyze(Q,p)
$.
\paragraph{\boldmath$Oracle(p)$} It returns the execution-based classification
$
c_o = Oracle(p)
$.
We implement $Oracle$ with sanitizers from major compiler toolchains, including AddressSanitizer, MemorySanitizer, UndefinedBehaviorSanitizer, and ThreadSanitizer. $P$ means that the relevant instrumented error was observed for the supplied input. While $N$ means only that no such error was observed in that execution, \sys treats this as proof for absence of error.

A query failure occurs whenever the two disagree, $c_q\neq c_o$. When $c_o=P\land c_q=N$, the query commits an \emph{observed false negative}; when $c_o=N\land c_q=P$, it commits an \emph{observed false positive} relative to the exercised execution harness.
We denote the tuple that documents this outcome as a \emph{witness} $w=(p,c_q,c_o)$. A witness $w$ also documents a query success, in which case $c_q=c_o$. A regression set $W_r$ stores every witness validated in earlier rounds, both successes and confirmed failures. Whenever a refined query is proposed, we recheck it against all the witnesses in $W_r$, so a fix for one weakness cannot silently break a case that was previously validated.

\paragraph{\boldmath$Gen(Q)$} %may generate multiple candidate programs. 
We ask the LLM to generate a pair of programs $p_0$ and $p_1$ that are as similar as possible while exposing contrasting query behavior. After validation, the pair forms a matched TN-FP or TP-FN witness pair. Near-identical programs encourage the LLM to reason about the relevant control or data flow pattern instead of matching incidental syntax such as function names. We request $Gen(Q)$ to alternate between TN-FP and TP-FN pairs because a refinement that improves observed recall may reduce observed precision, and vice versa. Failure to find a valid pair within the budget indicates only that no such weakness was found under the requested pair type and execution harness. Because $Gen(\cdot)$ is LLM-based, its proposed classifications are provisional, and a candidate is retained only after confirming $c_o=Oracle(p)\land c_q=Analyze(Q,p)$.

\begin{algorithm}[t]
\caption{ARQ query refinement loop.}
\label{alg:iteration}
\small
\Input{Initial query $Q_0$, iteration budget $T$, per-iteration budget $N$}
\Output{Refined query $Q_f$}
$Q\gets Q_0$; $W_r\gets\emptyset$\;
\BlankLine
\For{$t\gets1$ \KwTo $T$}{
    $W_c\gets\emptyset$\;
    \For{$i\gets1$ \KwTo $N$}{
        $W\gets\Gen(Q)$\label{alg:iteration:phase1}\;
        \If{$\forall(p,c_q,c_o)\in W:\ Analyze(Q,p)=c_q\land\Oracle(p)=c_o$}{
            $W_c\gets W$\; \textbf{break}
        }
    }
    \If{$W_c=\emptyset$}{
        \textbf{continue}\tcp{no valid witness found}\label{alg:iteration:noop}
    }
    $W_r\gets W_r\cup W_c$\; \textbf{break}
    \For{$j\gets1$ \KwTo $N$}{
        $Q'\gets\Improve(Q,W_c)$\label{alg:iteration:phase2}\;
        \If{$\forall(p,c_q,c_o)\in W_r:\ Analyze(Q',p)=c_o$}{
            $Q\gets Q'$\label{alg:iteration:success}\;
        }
    }
}
\Return $Q$\;
\end{algorithm}

\subsection{Refining the Query}
\label{s:design:refine}

After obtaining a validated witness set $W_c$, \sys uses it to guide a candidate query update. The candidate is checked against both $W_c$ and the regression set $W_r$ before it can replace the current query. If no candidate passes within the budget, the current query and regression set remain unchanged.

\paragraph{\boldmath$Improve(Q,W_c)$}
Given validated witnesses, it applies targeted edits to address the weaknesses exposed by $W_c$. Explicit witnesses focus the LLM-based improver on the observed failure rather than on an informal description of the query. It proposes a candidate query $Q'$.

The candidate is accepted only when it preserves every checked witness:
\begin{equation}
 \begin{aligned}
 \forall (p,c_q,c_o)\in W_c\cup W_r,Analyze(Q',p)=c_o. \nonumber
 \end{aligned}
\end{equation}
If this condition cannot be satisfied within the refinement budget, both $Q$ and the regression set $W_r$ remain unchanged.

In each iteration, the algorithm retries $Gen(Q)$ until it obtains a validated witness set or exhausts the budget. It then retries $Improve(Q,W_c)$ and retains the first candidate that passes $W_c\cup W_r$. Failed candidates are discarded, so exhausting the refinement budget leaves the current query unchanged.

The outer loop runs for at most $T$ iterations and adds newly validated witnesses to $W_r$ 
. Thus, the procedure provides regression protection over a finite, execution-grounded witness set, rather than a guarantee over all programs.

\subsection{System Implementation}
\label{s:design:impl}
We implement $Gen$ and $Improve$ with LLM-based coding agents equipped with task-specific tools. $Analyze$ invokes CodeQL on the generated program, while $Oracle$ invokes the relevant compiler sanitizer. The agents orchestrate these deterministic checks and use their results to guide the next action.

Likely due to limited training data~\cite{wang2026qlcoderquerysynthesizerstatic}, even modern LLMs can struggle with CodeQL syntax. During $Improve$, we provide the agent with the CodeQL Language Server through the Model Context Protocol~\cite{wang2026qlcoderquerysynthesizerstatic}. Both $Improve$ and $Analyze$ report syntax errors when they occur.

The budget $N$ can also be a time limit. This is useful for commercial agents where precise attempt limit is hard to enforce. With explicit budget control, the task in \cref{alg:iteration} is simple enough for agents backed by non-reasoning models.

\section{Evaluation}

In this section, we evaluate \sys to answer the following three research questions:

\squishlist
\item \textbf{RQ1:} How well do queries refined by \sys perform?
\item \textbf{RQ2:} Can \sys address real-world CodeQL issues and analyze real-world software?
\item \textbf{RQ3:} Does feedback from $Analyze$ and $Oracle$ improve refinement?
\squishend

We evaluate \sys using three mainstream commercial models as its underlying LLMs. Specifically, we use Gemini-3.5-flash in Google Antigravity~\cite{agyCli}, Claude-Sonnet-4.6 in Claude Code~\cite{claudeCli}, and GPT-5.4 in Codex~\cite{codexCli}. We configured each model's thinking level to be low or disabled to reduce cost. Throughout the remaining sections, we may abbreviate Gemini-3.5-flash as simply ``Gemini3.5'', Claude-Sonnet-4.6 as ``Sonnet4.6'', and GPT-5.4 as ``GPT5.4''.

\subsection{RQ1: Query Performance}

\label{s:rq1}
\paragraph{Dataset}
We evaluate these CodeQL queries on Juliet v1.3~\cite{julietv13,julietv11} and FormAI v2~\cite{formaiv2}. Juliet is a hand-crafted, labeled C/C++ dataset containing controlled vulnerability cases and flow variants, and it has been widely used to evaluate program analysis tools since its release~\cite{diversevul,castle,julietv13,formaiv2}. FormAI is a large corpus of C programs labeled by bounded model checkers.
We do not include other C/C++ datasets in the primary benchmark for two reasons. First, PrimeVul~\cite{primevul}, DiverseVul~\cite{diversevul}, and SECODEPLT~\cite{secodeplt} are primarily function- or file-level datasets rather than ready-to-run, project-scale benchmarks. PrimeVul provides CWE, CVE/NVD, repository, commit, and file metadata, and its labels are derived from vulnerability-fixing evidence rather than being simply LLM-generated. These metadata could be used to reconstruct vulnerable and patched project versions for a buildable subset. Doing so would require selecting repositories, resolving historical dependencies, reproducing the relevant builds and tests, and determining whether each recovered project exposes the exact query pattern targeted by a CodeQL query. That reconstruction process would introduce a new selection and buildability protocol, so we leave it for future work rather than treating the function-level samples as directly comparable to our project-scale runs. Castle~\cite{castle} is manually labeled but contains only 250 programs, too small for the scale of automated refinement and evaluation we require. Second, reliably compiling large numbers of real-world C/C++ projects is difficult given the variety of build systems and dependency managers in use. Juliet and FormAI provide controlled, reproducible compilation and labels for the present study, and we also evaluate the queries on seven real open-source projects in \cref{s:rq2}.

For Juliet v1.3, we only used samples targeting Linux. For FormAI v2, we discarded non-compilable samples and samples without provable classifications.

Our reusable evaluation artifact is the query-refinement task suite built around these benchmarks, rather than a new vulnerability dataset. It contains the 12 initial CodeQL queries, the refined-query outputs, execution-validated witness programs and contrastive pairs, regression cases, and the per-query detection results used to produce the aggregate tables. This artifact exposes the refinement inputs and outputs independently of \sys, allowing future query-maintenance methods to use the same tasks and compare against the same original queries. Juliet and FormAI remain evaluation substrates; their labels and synthetic or curated construction limit the conclusions that can be drawn about generalization.

\paragraph{Setup}
We evaluate 12 CodeQL queries on the two datasets described above. These queries cover common C/C++ vulnerability types, such as buffer-overflows, use-after-free, and double-free, and are drawn from the official CodeQL repository v2.23.1~\cite{codeqlgithub}, representative of the memory-safety and related CWE categories targeted by CodeQL's C/C++ query suite.

For each dataset, CodeQL is given 12 hours and 32 GiB memory to run each query on Ubuntu 24.04 on an AMD EPYC 7502 CPU.
Each query is refined for five iterations. We give each iteration a 25-minute wall-clock budget, shared by witness generation, query refinement, and validation.
\label{s:rq1:exc}

\paragraph{Baselines}
We compare against the original CodeQL queries. We do not include IRIS~\cite{li2024iris} or QLCoder~\cite{wang2026qlcoderquerysynthesizerstatic} because both target Java and address different tasks. IRIS infers sources and sinks for Java web-vulnerability analysis, whereas QLCoder synthesizes new Java CodeQL queries from CVE metadata. Neither provides a C/C++ implementation for refining an existing query with execution-validated witnesses. Porting either system to C/C++ and adapting its task definition would constitute a separate reimplementation study, so we leave that comparison to future work.

\paragraph{Metric}
As with most vulnerability detection tasks, we try to evaluate precision and recall. Query-level recall is difficult to interpret here because CodeQL queries and CWE labels are not in a one-to-one relationship. One query can be associated with several CWEs, while several queries can target different patterns under the same CWE. Moreover, a query may support only part of each CWE with which it is associated. An incorrect character conversion, for example, may also constitute an invalid pointer dereference, and each sample carries only one CWE label even though nothing prevents multiple weaknesses from occurring in the same program. We therefore report precision as the primary metric and additionally report \emph{query-averaged CWE-aligned coverage} in \cref{tab:rq1:res}.
\paragraph{Coverage definition}
For each query, we collect the set of vulnerable samples carrying any CWE associated with that query as $s_P$, and the corresponding set of true-positive samples detected by the query as $s_{TP}$. We compute $|s_P\cap s_{TP}|/|s_P|$ for each query and average across queries. A sample detected by multiple queries contributes once to each corresponding query--sample detection in the primary TP counts, but is deduplicated within each query's coverage calculation. This metric is a CWE-aligned coverage proxy, not exact recall for any individual query. Because the denominator can include patterns beyond a query's specific target, the absolute coverage values are expected to be low and should be interpreted comparatively rather than as complete vulnerability recall.
$
\text{Precision} = \frac{TP}{TP + FP}
$.
A \emph{true positive} (TP) is a query detection on a sample that contains a genuine instance of the targeted vulnerability, and a \emph{false positive} (FP) is a query detection on a sample that does not. 
TP and FP in the primary table are query--sample detections; the coverage column instead deduplicates samples within each query's associated CWE set. Thus, the two metrics answer different questions and should not be compared as if they had the same unit of analysis.

\paragraph{Results}
\begin{table}[t]
  \centering
  \small
  \setlength{\tabcolsep}{3pt}
  \resizebox{\columnwidth}{!}{%
  \begin{tabular}{l cccc | cccc}
    \toprule
    \multirow{2}{*}{System}
    & \multicolumn{4}{c}{Juliet v1.3}
    & \multicolumn{4}{c}{FormAI v2} \\
    \cmidrule(lr){2-5} \cmidrule(lr){6-9}
    & TP & FP & Prec.\,(\%) & Cov.\,(\%)
    & TP & FP & Prec.\,(\%) & Cov.\,(\%) \\
    \midrule
    CodeQL         & 5110 & 0  & 100 & 5.8 & 7009  & 96  & 98.6 & 1.7 \\
    Gemini3.5-\sys & 7957 & 14 & 99.8 & 6.3 & 15406 & 246 & 98.4 & 3.9 \\
    Sonnet4.6-\sys & 5489 & 1  & 100.0 & 8.7 & 12805 & 266 & 98.0 & 2.7 \\
    GPT5.4-\sys    & 4946 & 15 & 99.7 & 5.9 & 12457 & 195 & 98.5 & 1.8 \\
    \bottomrule
  \end{tabular}
  }
  \caption{
    TP, FP, and Precision are aggregated over query--sample detections. Cov. is query-averaged CWE-aligned coverage after deduplicating sample IDs within each query; it is not exact query-level recall. Per-query results are in appendices.
  }
  \label{tab:rq1:res}
\end{table}

\cref{tab:rq1:res} shows that \sys generally improves query performance, though the exact gains depend on the underlying model. On FormAI, all three models substantially increase true positive detection over CodeQL (\textbf{Gemini3.5: +119.8\%, Sonnet4.6: +82.7\%, GPT5.4: +77.7\%}). The coverage proxy increases from 1.7\% for CodeQL to 3.9\%, 2.7\%, and 1.8\% for Gemini3.5, Sonnet4.6, and GPT5.4, respectively. On Juliet, Gemini3.5 and Sonnet4.6 also improve over CodeQL (+55.7\% and +7.4\%, respectively), while \textbf{GPT5.4 shows a slight regression (-3.2\%)}. Coverage changes from 5.8\% for CodeQL to 6.3\%, 8.7\%, and 5.9\% for the three refined systems. Across all cases, the drop in Precision remains within 2.0\%. The refined queries therefore cover more evaluated cases, not all vulnerabilities in each CWE, because each CodeQL query targets only a subset of the patterns represented by its CWE labels. Additional efficiency measurements appear in the technical supplement.

Overall, the refinement loop improves coverage on the evaluated benchmarks. %The results do not establish complete recall or generalization beyond the evaluated witness and benchmark distributions; the 
GPT5.4--Juliet result shows that refinement may also regress.

\cref{tab:rq1:res} also shows why evaluating CodeQL queries is difficult. GPT5.4-\sys underperformed on Juliet but performed well on FormAI. A sample in the dataset may be labeled CWE-121 (Stack-Based Buffer Overflow), but \texttt{BadlyBoundedWrite.ql}, \texttt{NoSpaceForZeroTerminator.ql}, and \texttt{OverflowCalculated.ql} can all be seen as CWE-121. In addition, these queries may also be CWE-122 (Heap-Based Buffer Overflow). In contrast, each query only detects the specific type of vulnerability mentioned in its name. No existing datasets are labeled according to query type. They are mostly labeled according to CWE types, which is too coarse for CodeQL queries.

\subsection{RQ2: Real-World Impact}
\label{s:rq2}
To complement the curated benchmark evaluation, we construct a real-world evaluation set from unresolved CodeQL issue reproducers and widely used C/C++ projects. This setting tests whether the refined queries address problems encountered in practice rather than only patterns represented in synthetic or curated datasets.

\paragraph{Fixing GitHub CodeQL Issues}
The most direct application of \sys is fixing known, unresolved problems in queries described by open GitHub issues\footnote{\url{https://github.com/github/codeql/issues/}}. We selected the 3 most recent issues related to C/C++ CodeQL queries that have a minimally reproducible example, \#21187 (opened for 6 months), \#20577 (opened for 9 months), and \#16542 (opened for 27 months). All three happen to concern \texttt{UseAfterFree.ql}. We skipped the generation phase in \cref{alg:iteration} and directly used the examples supplied in each GitHub issue. \textbf{\sys, backed by each of the three models, produced a working fix for all three issues}, an improved version of \texttt{UseAfterFree.ql} that correctly classified (manually verified) the provided examples.

The minimally reproducible example incorporates a common code pattern where a pointer is passed through several functions before being used after it is freed.
The original \texttt{UseAfterFree.ql} misses this because it requires \lstinline[language=CodeQL]{defaultSourceSinkIsRelated}. \sys fixed this by extending control-flow tracking to function argument usage using \lstinline[language=CodeQL]{isNonThisParameterUse}. This produced a query that correctly flags the real bug reported in the issue.

\paragraph{Evaluating Queries on Open Source Projects}
We evaluated CodeQL queries and their refined counterparts on a real-world evaluation set of \textbf{seven widely used C/C++ projects} (libpng, zlib, curl, redis, pcre2, Mbed TLS, and TinyXML-2), spanning around 985K lines of code. The refined queries identified 2 previously unreported candidate bugs that the original CodeQL queries do not detect; we report their locations and triggering paths below. These findings are evidence of improved project-level coverage, not a measurement of project-wide precision or complete issue resolution. 

\texttt{InconsistentNullnessTesting.ql} found one candidate bug each in zlib\footnote{\url{https://github.com/madler/zlib/blob/da607da739fa6047df13e66a2af6b8bec7c2a498/adler32.c\#L70}} and libpng\footnote{\url{https://github.com/pnggroup/libpng/blob/d1d0abeffede1cc898ddc3d0e600839cf026d749/contrib/tools/pngfix.c\#L3936}}, two of the most widely used C libraries in the open-source ecosystem. The zlib finding was confirmed by maintainers, whereas the libpng finding remains a candidate. zlib implements the compression routines bundled in nearly every Linux distribution and used by tools such as Git and rsync (\textbf{7K} GitHub stars), while libpng is the reference implementation for reading and writing the PNG image format (\textbf{1.6K} GitHub stars). The zlib issue has been present since \textbf{2017}, while the libpng issue has been present since \textbf{October 2025}. Both findings concern cold code paths that are less frequently exercised and tested, which plausibly explains why they had not previously been caught.

\subsection{RQ3: Component Analysis}
\label{s:ablation}
We test whether comparing $Analyze$ with the execution result from $Oracle$ improves refinement. In Retrieval Augmented Generation, the grounding source is a knowledge database; in \sys, it is execution feedback.
\paragraph{Setup}
Procedures in \cref{s:rq1} are mimicked with one change. The agent no longer has access to $Analyze$ and $Oracle$ in \cref{alg:iteration}, so it cannot evaluate CodeQL queries on real programs, nor can it use the sanitizer. The agent still has access to the CodeQL Query Compiler and the LSP to reliably produce syntactically correct queries.

We use this approach as a \textit{baseline} because it relies on the agents' reasoning without execution feedback.

\paragraph{Results}
\begin{table}[t]
  \centering
  \small
  \setlength{\tabcolsep}{3pt}
  \begin{tabular}{l ccc | ccc}
    \toprule
    \multirow{2}{*}{Baseline}
    & \multicolumn{3}{c}{Juliet v1.3}
    & \multicolumn{3}{c}{FormAI v2} \\
    \cmidrule(lr){2-4} \cmidrule(lr){5-7}
    & TP & FP & Prec.\,(\%)
    & TP & FP & Prec.\,(\%) \\
    \midrule
    Gemini3.5 & 4970 & 0 & 100 & 7483 & 110 & 98.6 \\
    Sonnet4.6 & 2308 & 0 & 100 & 3146 & 40  & 98.7 \\
    GPT5.4    & 546  & 0 & 100 & 849  & 9   & 99.0 \\
    \bottomrule
  \end{tabular}
  \caption{
    Results for \textit{baseline}, aggregated over the same query subset as \cref{tab:rq1:res}. Per-query breakdown is in the technical supplement.
  }
  \label{tab:ablation:aggregate}
\end{table}

Comparing \cref{tab:ablation:aggregate} with \cref{tab:rq1:res}, \sys consistently outperforms this pure-LLM baseline. The baseline also regresses below vanilla CodeQL in \textbf{5 of 6} model/dataset configurations, compared to \textbf{1 of 6} for \sys (GPT5.4 on Juliet, \cref{s:rq1}). Manual inspection of agent transcripts shows why. Without execution feedback, the agent sometimes invents a weakness, such as a false positive in an already conservative query, and edits the query to ``fix'' it. The edit then removes true positives. This ablation does not isolate the separate contributions of $Analyze$, $Oracle$, matched pairing, or regression-witness accumulation; it measures their combined effect rather than showing that each component is individually necessary.

\section{Related Work}
\paragraph{LLM-assisted analysis and query synthesis.}
Recent systems combine LLMs with static analysis for vulnerability detection. IRIS~\cite{li2024iris} and Artemis~\cite{Ji_2025} use LLMs to identify taint sources and sinks, while QLCoder~\cite{wang2026qlcoderquerysynthesizerstatic} synthesizes Java CodeQL queries from CVE metadata. KNighter~\cite{Yang_2025} performs project-specific refinement using commit history, and MoCQ~\cite{li2025automatedstaticvulnerabilitydetection} uses symbolic validation for general bug patterns. These approaches differ from ARQ in their language, input, or objective. QLCoder generates new Java queries from external metadata; ARQ instead refines an existing C/C++ query with independently synthesized, execution-validated programs. This setting targets memory-lifetime and pointer-aliasing behavior that is not directly captured by conventional taint labels.

\paragraph{Execution-grounded agents and benchmarks.}
LLM agents have also been connected to fuzzers, language servers, and automated repair tools~\cite{yang2024sweagentagentcomputerinterfacesenable,codamosa,lspai,Blinn_2024,aprLLM0,joshi2022repairnearlygenerationmultilingual}. ARQ uses these tools to generate contrastive witnesses and validate query updates with compiler sanitizers. %SECODEPLT~\cite{secodeplt} provides a benchmark of dynamically evaluated vulnerable and patched C/C++ artifacts, tests, and proof-of-concept evidence. We leave evaluation on SECODEPLT to future work.

\section{Conclusions}
We presented \sys, an agentic framework for automatically refining static
analysis queries through execution-validated program synthesis and iterative
LLM-guided reasoning. Across 12 CodeQL queries on Juliet and FormAI, \sys
substantially improves true positive detection while keeping Precision at least
98.0\%. Beyond these benchmarks, \sys addressed 3 previously unresolved GitHub
issues in the CodeQL repository and identified 2 previously unreported
candidate bugs in 7 widely used C/C++ projects. The results show that
execution-grounded, example-driven query refinement can improve coverage on
the evaluated witnesses, benchmarks, and projects, while leaving complete
recall and project-wide precision as open evaluation questions.

\section*{Limitations}
\sys's performance depends directly on the quality of $Oracle$. Even if $Oracle$ returns $c_o=N$, this does not prove that the program is vulnerability-free; it means only that no relevant error was observed for the supplied input and execution path. A vulnerable path may not have been exercised. Compiler sanitizers can provide strong evidence when they report an error, but a non-report does not establish global safety.

Because CodeQL query patterns do not map one-to-one to dataset-level CWE labels, query-averaged CWE-aligned coverage is only a best-effort estimation of recall and may not faithfully reflect the actual recall of a specialized query.

Our ablation also combines several execution-grounded components, and the real-world study reports candidate findings and issue reproductions rather than project-wide precision or complete issue resolution. Future work should evaluate the method on independently constructed project benchmarks and with component-level ablations.

\bibliography{main}
\clearpage
\clearpage
\onecolumn

\appendix
\setcounter{secnumdepth}{2}
\renewcommand{\thesection}{Appendix \Alph{section}.}

\begin{multicols}{2}

\section{Raw Per-Query Results}
\label{app:raw}

We report the per-query results underlying the aggregate evaluation in
Section~4.1. Each row corresponds to one official CodeQL query, and each
pair of columns reports the number of true-positive (TP) and false-positive
(FP) query--sample detections produced by the original CodeQL query, the
LLM-only baseline, and the corresponding \sys-refined query for each model.
Tables~\ref{tab:raw-results}(a) and~\ref{tab:raw-results}(b) show results
on Juliet v1.3 and FormAI v2, respectively.

Because the same sample may be detected by multiple queries, summing the
TP and FP counts across rows do not equal to the aggregate results in
Tables~1--2, which use the evaluation procedure described in Section~4.1.

"--" indicates that no successfully refined query was produced or that
evaluation did not complete within the 12-hour limit. These cases are
concentrated in queries that rely on computationally expensive path-sensitive,
interprocedural, or range-based analyses. For example,
\texttt{OverrunWriteProductFlow.ql} combines product-flow analysis, global
data flow, and range analysis to jointly track buffer and size expressions,
while \texttt{InvalidPointerDeref.ql} reasons about pointer dereferences
relative to allocation bounds. Such queries are more costly to repeatedly modify and re-evaluate during iterative refinement, increasing the likelihood of exhausting the evaluation budget.

\end{multicols}

\vspace{10mm}

\begin{center}

\fontsize{6.5}{7.0}\selectfont
\setlength{\tabcolsep}{2pt}
\renewcommand{\arraystretch}{0.75}

\resizebox{0.98\textwidth}{!}{%
\begin{tabular}{l cc cc cc cc cc cc cc}
\toprule
\multirow{2}{*}{Query}
& \multicolumn{2}{c}{CodeQL}
& \multicolumn{2}{c}{Gemini3.5-Base}
& \multicolumn{2}{c}{Gemini3.5-\sys}
& \multicolumn{2}{c}{Sonnet4.6-Base}
& \multicolumn{2}{c}{Sonnet4.6-\sys}
& \multicolumn{2}{c}{GPT5.4-Base}
& \multicolumn{2}{c}{GPT5.4-\sys} \\

\cmidrule(lr){2-3}
\cmidrule(lr){4-5}
\cmidrule(lr){6-7}
\cmidrule(lr){8-9}
\cmidrule(lr){10-11}
\cmidrule(lr){12-13}
\cmidrule(lr){14-15}

& TP & FP
& TP & FP
& TP & FP
& TP & FP
& TP & FP
& TP & FP
& TP & FP \\
\midrule

\texttt{MemoryMayNotBeFreed.ql}         & 4512 & 0 & 4343 & 0 & 5047 & 0  & 1794 & 0 & 4497 & 0 & 0   & 0 & 4180 & 0\\
\texttt{WcharCharConversion.ql}         & 0    & 0 & 0    & 0 & 0    & 0  & 0    & 0 & 1    & 0 & 0   & 0 & 0    & 0\\
\texttt{BadlyBoundedWrite.ql}           & 156  & 0 & 156  & 0 & 216  & 0  & 156  & 0 & 228  & 0 & 156 & 0 & 240  & 0\\
\texttt{NoSpaceForZeroTerminator.ql}    & 0    & 0 & 0    & 0 & 0    & 0  & 0    & 0 & 0    & 0 & 0   & 0 & 0    & 0\\
\texttt{OverflowCalculated.ql}          & 0    & 0 & 0    & 0 & 0    & 0  & 0    & 0 & 0    & 0 & 0   & 0 & 77   & 15\\
\texttt{DoubleFree.ql}                  & 153  & 0 & 243  & 0 & 243  & 0  & 34   & 0 & 659  & 1 & 0   & 0 & 187  & 0\\
\texttt{OverrunWriteProductFlow.ql}     & 0    & 0 & 0    & 0 & 1810 & 0  & 0    & 0 & 0    & 0 & 0   & 0 & 0    & 0\\
\texttt{InconsistentNullnessTesting.ql} & 38   & 0 & 38   & 0 & 38   & 0  & 38   & 0 & 39   & 0 & 38  & 0 & 38   & 0\\
\texttt{NewArrayDeleteMismatch.ql}      & 187  & 0 & 211  & 0 & 343  & 0  & 187  & 0 & 199  & 0 & 199 & 0 & 199  & 0\\
\texttt{SizeCheck.ql}                   & 0    & 0 & 0    & 0 & 73   & 0  & 0    & 0 & 0    & 0 & 0   & 0 & 0    & 0\\
\texttt{UseAfterFree.ql}                & 153  & 0 & 153  & 0 & 0    & 0  & 153  & 0 & 153  & 0 & 153 & 0 & 153  & 0\\
\texttt{InvalidPointerDeref.ql}         & 0    & 0 & 0    & 0 & 702  & 14 & 0    & 0 & 0    & 0 & 0   & 0 & --   & --\\

\bottomrule
\end{tabular}
}

\vspace{1mm}
{\small\textbf{(a) Juliet v1.3}}

\vspace{2mm}

\resizebox{0.98\textwidth}{!}{%
\begin{tabular}{l cc cc cc cc cc cc cc}
\toprule
\multirow{2}{*}{Query}
& \multicolumn{2}{c}{CodeQL}
& \multicolumn{2}{c}{Gemini3.5-Base}
& \multicolumn{2}{c}{Gemini3.5-\sys}
& \multicolumn{2}{c}{Sonnet4.6-Base}
& \multicolumn{2}{c}{Sonnet4.6-\sys}
& \multicolumn{2}{c}{GPT5.4-Base}
& \multicolumn{2}{c}{GPT5.4-\sys} \\

\cmidrule(lr){2-3}
\cmidrule(lr){4-5}
\cmidrule(lr){6-7}
\cmidrule(lr){8-9}
\cmidrule(lr){10-11}
\cmidrule(lr){12-13}
\cmidrule(lr){14-15}

& TP & FP
& TP & FP
& TP & FP
& TP & FP
& TP & FP
& TP & FP
& TP & FP \\
\midrule

\texttt{MemoryMayNotBeFreed.ql}         & 6127 & 81 & 6127 & 81 & 11445 & 137 & 6127 & 81 & 11636 & 236 & 0   & 0 & 6381 & 77\\
\texttt{WcharCharConversion.ql}         & 0    & 0  & 0    & 0  & 0     & 0   & 0    & 0  & 0     & 0   & 0   & 0 & 0    & 0\\
\texttt{BadlyBoundedWrite.ql}           & 47   & 1  & 47   & 1  & --    & --  & 47   & 1  & 51    & 1   & 47  & 1 & 64   & 1\\
\texttt{NoSpaceForZeroTerminator.ql}    & 159  & 1  & 485  & 0  & 384   & 16  & 485  & 6  & 166   & 1   & 159 & 1 & 175  & 1\\
\texttt{OverflowCalculated.ql}          & 112  & 7  & 80   & 7  & 21    & 1   & 80   & 7  & 125   & 7   & 0   & 0 & 5745 & 109\\
\texttt{DoubleFree.ql}                  & 41   & 0  & 98   & 11 & 3025  & 95  & 98   & 11 & 789   & 19  & 0   & 0 & 106  & 1\\
\texttt{OverrunWriteProductFlow.ql}     & --   & -- & --   & -- & --    & --  & --   & -- & --    & --  & --  & --& --   & --\\
\texttt{InconsistentNullnessTesting.ql} & 84   & 1  & 84   & 1  & 518   & 7   & 84   & 1  & 518   & 7   & 84  & 1 & 245  & 7\\
\texttt{NewArrayDeleteMismatch.ql}      & 0    & 0  & 0    & 0  & 0     & 0   & 0    & 0  & 0     & 0   & 0   & 0 & 0    & 0\\
\texttt{SizeCheck.ql}                   & 214  & 3  & 388  & 3  & 469   & 10  & 222  & 3  & 388   & 3   & 214 & 3 & 219  & 3\\
\texttt{UseAfterFree.ql}                & 350  & 3  & 350  & 3  & 831   & 17  & 350  & 3  & 362   & 3   & 350 & 3 & 362  & 3\\
\texttt{InvalidPointerDeref.ql}         & --   & -- & --   & -- & --    & --  & --   & -- & --    & --  & --  & --& --   & --\\

\bottomrule
\end{tabular}
}

\vspace{1mm}
{\small\textbf{(b) FormAI v2}}

\vspace{1.5mm}

\captionof{table}{
Per-query TP/FP results for the evaluated CodeQL queries.
}
\label{tab:raw-results}

\end{center}

\twocolumn

\end{document}